\documentclass[12pt,draftcls,journal,onecolumn]{IEEEtran}

\usepackage[T1]{fontenc}
\usepackage[latin9]{inputenc}
\usepackage{color}
\usepackage{graphicx}
\usepackage{epsfig}
\usepackage{amssymb}

\usepackage{amsmath}

\newtheorem{example}{Example}
\newtheorem{definition}{Definition}

\newtheorem{lemma}{Lemma}

\begin{document}

\title{ML Decoding Complexity Reduction in STBCs Using Time-Orthogonal Pulse Shaping}

 \author{
\IEEEauthorblockN{Rakshith Rajashekar and K.V.S. Hari,~\IEEEmembership{Senior Member,~IEEE}\\ 
        \IEEEauthorblockA{Department of Electrical Communication Engineering,\\ 
      Indian Institute of Science, Bangalore 560012\\ 
      \{rakshithmr, hari\}@ece.iisc.ernet.in}
}}

\maketitle

\begin{abstract}
Motivated by the recent developments in the Space Shift Keying (SSK) and Spatial Modulation (SM) systems which employ Time-Orthogonal Pulse Shaping (TOPS) filters to achieve transmit diversity gains, we propose TOPS for Space-Time Block Codes (STBC). We show that any STBC whose set of weight matrices partitions into $P$ subsets under the equivalence relation termed as  {\em Common Support Relation} can be made $P$-group decodable by  properly employing TOPS waveforms across space and time. Furthermore, by considering some of the well known STBCs in the literature we show that the {\em order} of their Maximum Likelihood decoding complexity can be greatly reduced by the application of TOPS.
\end{abstract}


\section{Introduction}

It is well known that the multi-antenna transmission scheme is imperative for reliable, high data rate, wireless communication which unfortunately comes at the cost of substantial decoding complexity at the receiver. The inherent cause for high decoding complexity at the receiver is the Inter-Antenna Element (IAE) interference which arises due to the simultaneous activation of all the antennas at the transmitter. For example, in the Vertical Bell Laboratories Layered Space-Time (V-BLAST) architecture \cite{VBLAST}, the Maximum Likelihood (ML) decoding complexity is exponential in the number of transmit antennas. Though, the Sphere Decoding (SD) \cite{SD1} algorithm with sufficiently large search-radius guarantees ML performance \cite{SD2}, \cite{SD3} the reduction in the search complexity due to SD at low Signal-to-Noise Ratios (SNR) is very low compared to that at the high SNRs. Furthermore, the V-BLAST scheme offers the maximum multiplexing gain while the diversity gain offered by it is only one. In order to achieve full transmit diversity, Space-Time Block Codes (STBC) \cite{TSC} were developed while aiming at reducing the decoding complexity as much as possible. Some examples of these codes whose decoding complexities are same as that of a Single-Input Multiple-Output (SIMO) system are Alamouti code \cite{ALAM}, STBCs from orthogonal designs \cite{OD}, and codes from Complex Interleaved Orthogonal Design (CIOD) \cite{SSDCIOD}. Though these codes offer low complexity ML decoding, their achievable multiplexing gains are limited. In \cite{CDA1}, full diversity, maximum multiplexing gain codes for arbitrary number of transmit antennas were proposed which also have been shown to have Non Vanishing Determinant (NVD) property under certain conditions \cite{VITERBO}. Current research in the area of STBCs is focused on obtaining codes having full diversity, high multiplexing gain, NVD property, and low ML decoding complexity. Reference \cite{SRCODE} gives a comparison of the decoding complexities of various best known STBCs for $2\times2$ and $4\times2$ Multiple-Input Multiple-Output (MIMO) systems. Some recent results in this area for some asymmetric MIMO configurations such as $4\times2$, $6\times2$, etc, can be found in \cite{FAST1}, \cite{FAST2}. 

Spatial Modulation (SM) \cite{SM1}, \cite{SM2} and Space Shift Keying (SSK) \cite{SSK1} are some of the novel developments in the family of low complexity MIMO schemes that reduce the ML decoding complexity at the receiver by completely removing the IAE interference. Specifically, in these schemes only one transmit antenna is activated in any symbol duration which results in the complete removal of IAE interference, but this results in the transmit diversity order of only one and also limits the achievable rates. In \cite{TOSD-SM}, it was shown that a transmit diversity order of more than one can be achieved in SM systems by employing Time-Orthogonal Pulse Shaping (TOPS) filters. In \cite{TOSD-SM-STBC}, as an improvement over \cite{TOSD-SM}, the conventional SM scheme is extended by activating two transmit antennas for the transmission of the Alamouti code \cite{ALAM} and by employing TOPS waveforms over partitioned antenna pairs to achieve a transmit diversity order of two. Recently, the authors in \cite{SSK-TOSD} have compared the bandwidth requirement of the conventional SM scheme assumed to be employing raised-cosine filter/half-sine filter with that of  the TOPS assisted SSK scheme employing Hermite polynomial based TOPS filter \cite{HERM}. It was shown in \cite{SSK-TOSD} that under stringent conditions on the energy containment, the bandwidth required by the TOPS assisted SM scheme is lesser than that of the conventional SM scheme.

Against this background following are the novel contributions of this paper:

$\bullet$ Inspired by the application of TOPS filter in SM/SSK scheme, we propose TOPS for STBCs. We show that by properly applying the TOPS waveforms across space and time the IAE interference can be partially removed thereby reducing the ML decoding complexity of the STBCs. Specifically, given a design of an STBC we define an equivalence relation on its set of weight matrices and partition the set into disjoint subsets which do not overlap in space and time. If the number of such disjoint subsets is $P$, then $P$ distinct TOPS waveforms are assigned to $P$ subsets. The transmitted symbols belonging to different subsets become separable over space and time since they do not overlap in space and time and also their associated shaping waveforms are time-orthogonal. Thus, an STBC having $P$ disjoint subsets which don not overlap in space and time becomes $P$-group decodable.

$\bullet$ We study the reduction in the ML decoding complexity by the proposed application of the TOPS waveforms in some of the best known STBCs in the literature. We show that the order of the ML decoding complexity in these STBCs can be greatly reduced due to the application of the TOPS filters.

The rest of the paper is organized as follows. Section II describes the system model considering distinct pulse shaping filters across transmit antennas and channel uses. Section III describes the proposed application of TOPS filters in STBCs and gives a systematic method for reducing the ML decoding complexity in STBCs. In Section IV, we study the reduction in the decoding complexities considering some of the best known codes in the literature. Finally, Section V concludes the paper.

{\em Notation:} Boldface uppercase letters such as $\mathbf H$ represent matrices. Boldface lowercase letters such as $\mathbf y$ represent vectors. The operator $\| \cdot\|_2$ represents the two-norm of a vector, $\|\cdot \|_F$ represents the Frobenius norm of a matrix, and the operator $\langle\cdot,\cdot\rangle$ represents the inner product of two given vectors or the correlation of two given waveforms. The subscripts I and Q, as in $s_I$ and $s_Q$, are used to represent the real and imaginary components of a complex number. The $|\cdot|$ is used to represent the magnitude of a complex quantity, or the cardinality of a given set.
Blackboard-bold font letters $\mathbb R$ and $\mathbb C$ represent the field of real and imaginary numbers, respectively.

\section{System model}

Consider a MIMO system having $N_t$ transmit and $N_r$ receive antennas and a quasi static Rayleigh fading channel as follows:
\begin{equation}
\footnotesize
 \underbrace{\left[ {\mathbf y}_1(t),{\mathbf y}_2(t),\hdots,{\mathbf y}_T(t)\right]}_{{\mathbf Y}(t)} = \underbrace{\left[ {\mathbf h}_1,{\mathbf h}_2,\hdots,{\mathbf h}_{N_t}\right]}_{\mathbf H}
\underbrace{\left[\begin{array}{cccc}
           w_{11}(t)x_{11} & w_{12}(t)x_{12} & \cdots & w_{1T}(t)x_{1T}\\
	  w_{21}(t)x_{21} & w_{22}(t)x_{22} & \cdots & w_{2T}(t)x_{2T}\\
	  \vdots & \vdots & \ddots & \vdots\\
          w_{N_t1}(t)x_{N_t1} & w_{N_t2}(t)x_{N_t2} & \cdots & w_{N_tT}(t)x_{N_tT}
\end{array}\right]}_{{\mathbf X}(t)} + \underbrace{\left[ {\mathbf n}_1(t),{\mathbf n}_2(t),\hdots,{\mathbf n}_T(t)\right]}_{{\mathbf N}(t)},
\label{STSYSMODEL}
\end{equation}
where, $x_{ij}$ represents the $(i,j)^{\text{th}}$ element of the transmitted space-time matrix ${\mathbf X}\in {\mathbb C}^{N_t\times T}$, the channel matrix ${\mathbf H}\in {\mathbb C}^{N_r\times N_t}$ whose elements are from ${\cal CN}(0,1)$ is assumed to remain constant over $T$ channel uses, ${\mathbf n}_i(t)$ for any $1\leq i\leq T$ represents the complex AWGN with independent real and imaginary components having identical autocorrelation function $R_n(\tau)=\frac{N_0}{2}\delta(\tau)$, and ${\mathbf y}_i(t)=\sum_{j=1}^{N_t}w_{ji}(t)x_{ji}{\mathbf h}_j+{\mathbf n}_i(t)$ for any $1\leq i \leq T$ represents the received waveform at the $i^{\text{th}}$ channel use across $N_r$ receive antennas, where $w_{ji}(t)$ for $t=[0,T_s]$ is the pulse shaping waveform used by the $j^{\text{th}}$ transmit antenna at the $i^{\text{th}}$ channel use and $T_s$ represents the symbol duration. In this paper we consider only minimum delay codes, that is, codes having $T=N_t$.

Note that in the conventional MIMO system we have $w_{i,j}(t)=w(t)$ for all legitimate pairs $(i,j)$ where $w(t)$ is the raised-cosine / half-sine filter \cite{RACOS} in which case the system in Eq.(\ref{STSYSMODEL}) reduces after matched filtering to ${\mathbf Y}={\mathbf H}{\mathbf X}+{\mathbf N}$, where the $(i,j)^{\text{th}}$ element of ${\mathbf N}$ is given by $\int_0^{T_s}w(t)n_{ij}(t)dt$. Considering perfect Channel State Information at the Receiver (CSIR) the ML solution is given by
\begin{equation}
 \hat{\mathbf X}_{ML}=\arg \min_{\mathbf X \in {\cal C}} \| {\mathbf Y}-{\mathbf H}{\mathbf X}\|_F^2,
\end{equation}
where, $\cal C$ represents the codebook employed at the transmitter.

In order to achieve full diversity we need to have $rank({\mathbf X}_1-{\mathbf X}_2)=N_t$ for all ${\mathbf X}_1\neq{\mathbf X}_2 \in \cal C$, and the coding gain defined by 
\begin{equation}
 G=\min_{{\mathbf X}_1\neq{\mathbf X}_2 \in \cal C}\det{\left[({\mathbf X}_1-{\mathbf X}_2)({\mathbf X}_1-{\mathbf X}_2)^H\right]} 
\end{equation}
should be maximized to achieve low Bit Error Rate (BER) performance \cite{TSC}.

\section{Proposed application of TOPS filters in STBCs}

Any linear STBC encoding $K$ real information symbols can be written in terms of weight matrices as 
\begin{equation}
 {\mathbf X}=\sum_{i=1}^{K}s_i{\mathbf A}_i,
\label{LSTBC}
\end{equation}
where, $s_i$ are assumed to be from a PAM signal set. Let the set containing all the weight matrices ${\mathbf A}_i\in {\mathbb C}^{N_t\times N_t}$ be represented by $\cal A$.

\begin{definition}
 Support Set (SS) of a matrix $\mathbf B\in {\mathbb C}^{m\times n}$ is defined as 
\begin{equation}
SS(\mathbf B)=\{ (i,j)~|~{\mathbf B}(i,j)\neq 0 , 1\leq i\leq m, 1\leq j\leq n\}.
\end{equation}
\end{definition}

\begin{definition} 
 For any two weight matrices ${\mathbf A}_k,{\mathbf A}_l \in {\cal A}$, we will have either $SS({\mathbf A}_k)=SS({\mathbf A}_l)$, or $SS({\mathbf A}_k)\neq SS({\mathbf A}_l)$. Thus, we have a relation say $\sim$ on the set $\cal A$ which we term as the Common Support Relation (CSR).
\end{definition}

\begin{lemma}
 The CSR $\sim$ on the set $\cal A$ is an equivalence relation, and hence, partitions $\cal A$ into disjoint subsets given by ${\cal A}_1$, ${\cal A}_2$ , $\hdots$ , ${\cal A}_P$, with $P\geq1$.
\end{lemma}
{\em Proof:} Straightforward.

Let $|{\cal A}_1|=g_1$, $|{\cal A}_2|=g_2$ , $\hdots$ , $|{\cal A}_P|=g_P$, which yields $\sum_{i=1}^{P}g_i=K$, and let the corresponding index set be ${\cal I}_i=\{j~|~1\leq j\leq K~ \ni {\mathbf A}_j\in {\cal A}_i\}$. Thus, we can write $\mathbf X$ in Eq.(\ref{LSTBC}) as a sum of non-overlapping (in space and time) set of matrices as follows:
\begin{equation}
 {\mathbf X}=\sum_{i=1}^{P}{\mathbf X}_i,~\text{where}~{\mathbf X}_i=\sum_{k\in {\cal I}_i}s_k{\mathbf A}_k.
\end{equation}
We restrict all the symbols belonging to a given partition ${\mathbf X}_i$ to have their associated shaping filters to be same $w_i(t)$ and further the shaping filters associated with different partitions are chosen to be time-orthogonal. That is, $P$ time-orthogonal pulses of duration $T_s$ are obtained using \cite{HERM} which occupy the same bandwidth, and are assigned to $P$ distinct partitions, which gives
\begin{equation}
 {\mathbf X}(t)=\sum_{i=1}^{P}{\mathbf X}_iw_i(t),
\end{equation}
and hence,
\begin{equation}
 {\mathbf Y}(t)=\sum_{i=1}^{P}{\mathbf H}{\mathbf X}_iw_i(t)+{\mathbf N}(t).
\end{equation}

It is easy to see that after matched filtering across $N_r$ receive antennas and over $T$ time slots, we arrive at
\begin{equation}
 {\mathbf Y}_i={\mathbf H}{\mathbf X}_i + {\mathbf N}_i,
\end{equation}
for $1\leq i \leq P$, where the $(i,j)^{\text{th}}$ element of ${\mathbf N}_i$ is given by $\int_0^{T_s}w_i(t)n_{ij}(t)dt$. Thus, the STBC becomes $P$-group decodable with the aid of $P$ time-orthogonal waveforms. The ML solution in this case is given by 
\begin{equation}
 \hat{\mathbf X}_{ML}=\sum_{i=1}^{P}\hat{\mathbf X}_{i~ML}, ~\text{where} ~\hat{\mathbf X}_{i~ML}=\arg \min_{\text{over all} ~ \mathbf{X}_i}\|{\mathbf Y}_i-{\mathbf H}{\mathbf X}_i\|_F^2.
\end{equation}

Further reduction in the ML decoding complexity may be possible if the sub-codes ${\mathbf X}_i$ are themselves group decodable, i.e. if each of the ${\cal A}_i$ can be partitioned into disjoint subsets ${\cal A}_1^{i},{\cal A}_2^{i},\hdots,{\cal A}_Q^{i}$ such that ${\mathbf A}_k{\mathbf A}_l^H+{\mathbf A}_l{\mathbf A}_k^H={\mathbf 0}~\forall~ {\mathbf A}_k\in {\cal A}_m^{i}$, ${\mathbf A}_l\in {\cal A}_n^{i}$, and $1\leq m\neq n\leq Q$.

Note that with the increase in $P$ for a fixed $T_s$ the bandwidth requirement of the TOPS filters increases \cite{HERM}. However, for $P=4$  it was shown in \cite{SSK-TOSD} that the bandwidth requirement of the TOPS filters designed for narrow band communication is less than that of the raised-cosine / half-sine filters. We see in the next section that $P=2$ is sufficient in reducing the ML decoding complexities of some of the best known codes designed for $2\times 2$, $4\times 2$ MIMO systems.

\section{ML decoding complexity reduction in the existing STBCs}

In this section we study the reduction in the order of the ML decoding complexity due to TOPS filters considering some examples.

\begin{example}[\textbf{V-BLAST}]
 Consider a V-BLAST system having $N_t=4$ and employing a square-QAM signal set of size $M$. The ML decoding complexity of this scheme is $M^4$. In this scheme we have $K=8$ and $T=1$. It is easy to verify from Eq.(\ref{LSTBC}) that $\cal A$ partitions into $4$ subsets. Employing four time-orthogonal pulse waveforms across four transmit antennas enables single symbol ML decoding at the receiver which reduces the decoding complexity to $M$. The employment of separable QAM signal set enables real and imaginary parts to be decoded independently, which further reduces the decoding complexity to $\sqrt{M}$. Furthermore, by hard-limiting (quantizing) the real and imaginary parts we can achieve a decoding complexity that is independent of the constellation size.
\end{example}

\subsection{Codes for $2\times2$ MIMO system}
The best known codes in the literature for a $2\times 2$ MIMO system are 
\begin{itemize}
 \item The Golden code \cite{THEGOLD}, and
 \item The $2\times 2$ Srinath-Rajan Code \cite{SRCODE}.
\end{itemize}
Both these codes have the NVD property, identical coding gain and ML decoding complexity of $M^2\sqrt{M}$ when a separable QAM signal set of size $M$ is used.

\begin{example}[\textbf{Golden code}]
 The Golden code of \cite{THEGOLD} is given by
\begin{equation}
 \frac{1}{\sqrt{5}}\left[\begin{array}{cc}
                         \alpha(a+b\theta) & \alpha(c+d\theta)\\
			 \gamma {\bar\alpha}(c+d\bar\theta) & {\bar\alpha}(a+b\bar\theta)\\
                        \end{array}
\right]
\end{equation}
where, $a,b,c,d\in \mathbb{Z}[i]$, $\theta=\frac{1+\sqrt{5}}{2}$, $\bar{\theta}=1-\theta$, $\alpha=1+i-i\theta$, $\bar\alpha=1+i(1-\bar{\theta})$ and $\gamma \in {\mathbb C}$. In this case $\cal A$ partitions into $2$ subsets, and we will have $w_{11}(t)=w_{22}(t)=w_0(t)$ and $w_{12}(t)=w_{21}(t)=w_1(t)$, where $w_0(t)$ and $w_1(t)$ are the two distinct time-orthogonal shaping filters. Since, the pulse shaping waveforms across diagonal and off-diagonal elements are time-orthogonal with each other, the symbols $a$ and $b$ can be decoded independently of the symbols $c$ and $d$. Thus, the decoding complexity reduces to $M^2$. Since, the real and imaginary components can be decoded independently the decoding complexity further reduces to $M$. With the aid of the QR decomposition and hard-limiting, decoding complexity can be further reduced to $\sqrt{M}$.
\end{example}

\begin{example}[\textbf{Srinath-Rajan code}]
 The $2\times 2$ Srinath-Rajan code of \cite{SRCODE} is given by
\begin{equation}
 \left[\begin{array}{cc}
        s_{1I}+is_{2Q} & \sqrt{i}(s_{3I}+is_{4Q})\\
	\sqrt{i}(s_{4I}+is_{3Q}) & s_{2I}+is_{1Q}
       \end{array}
\right],
\end{equation}
where $s_{iI}=x_{iI}\cos\theta-x_{iQ}\sin\theta$ and $s_{iQ}=x_{iI}\sin\theta+x_{iQ}\cos\theta$ for $i=1,2,3,4$, $\theta=\frac{\arctan{2}}{2}$, and $x_{1I},x_{1Q},x_{2I},x_{2Q}\in {\mathbb Z}$. As in the case of the Golden code this code also partitions into two subsets, one set includes the symbols along the diagonal and the other includes the off-diagonal symbols. Since, each of the sub-codes is a CIOD  in itself, they are single-symbol decodable. Thus the complexity reduces to $M$. Furthermore, with the aid of the QR decomposition and hard-limiting the ML decoding complexity can be further reduced to $\sqrt{M}$ which is same as that of the Golden code.
\end{example}

\subsection{Codes for $4\times2$ MIMO system}

The best known codes in the literature for a $4\times 2$ MIMO system are 
\begin{itemize}
\item The $4\times 2$ Srinath-Rajan Code \cite{SRCODE}, and
\item Fast decodable $4\times 2$ code of  \cite{FAST2}.
 \end{itemize}
The ML decoding complexity of the $4\times 2$ Srinath-Rajan code was shown to be $M^4\sqrt{M}$ \cite{SRCODE} when a separable QAM of size $M$ is used. On the other hand the ML decoding complexity of the Fast decodable $4\times 2$ code was shown to be $M^5$ \cite{FAST2}.

\begin{example}[\textbf{$4\times 2$ Srinath-Rajan code}]
The $4\times 2$ Srinath-Rajan code of \cite{SRCODE} is given by
\begin{equation}
 \left[\begin{array}{cccc}
        s_{1I}+is_{3Q} & -s_{2I}+is_{4Q} & \sqrt{i}(s_{5I}+is_{7Q}) & \sqrt{i}(-s_{6I}+is_{8Q})\\
	s_{2I}+is_{4Q} &  s_{1I}-is_{3Q} & \sqrt{i}(s_{6I}+is_{8Q}) & \sqrt{i}(s_{5I}-is_{7Q})\\
	\sqrt{i}(s_{7I}+is_{5Q}) & \sqrt{i}(-s_{8I}+is_{6Q}) & s_{3I}+is_{1Q} & -s_{4I}+is_{2Q}\\
	\sqrt{i}(s_{8I}+is_{6Q}) & \sqrt{i}(s_{7I}-is_{5Q}) & s_{4I}+is_{2Q} & s_{3I}-is_{1Q}
       \end{array}
\right],
\end{equation}
 where, $s_{iI}$ and $s_{iQ}$ are same as that given in Example 3 for $i=1,2,\hdots,8$. It is easy to see that $\cal A$ partitions into two subsets where one includes the $2\times2$ diagonal blocks and the other includes the $2\times2$ off-diagonal blocks each of which can be seen as a CIOD for four transmit antennas. Thus, the decoding complexity reduces to $M$ as the CIOD is single symbol decodable, with the aid of the QR decomposition and hard-limiting the decoding complexity can be further reduced to $\sqrt{M}$.
\end{example}

\begin{example}[\textbf{Fast decodable $4\times 2$ code}]
The $4\times 2$ Fast decodable code of \cite{FAST2} is given by
\begin{equation}
 \left[\begin{array}{cccc}
        s_{1} & -r^2s_1^* & -r^3\sigma(s_4) & -r\sigma(s_3)^*\\
	r^2s_{2} & s_1^* &  r\sigma(s_3) & -r^3\sigma(s_4)^*\\
	rs_{3} & -r^3s_3^* & \sigma(s_1) & -r^2\sigma(s_2)^*\\
	r^3s_{3} & rs_2^* & r^2\sigma(s_1) & \sigma(s_1)^*
       \end{array}
\right],
\end{equation} 
where, $s_i=f_{4i-3}u1+f_{4i-2}u2+f_{4i-1}u3+f_{4i}u4$, $f_i$ are from a PAM signal set and the $\mathbb Z$-bases $\{u1,u2,u3,u4\}=\{1,\zeta+\zeta^{-1},\frac{\zeta-\zeta^{-1}}{2},\frac{\zeta^2-\zeta^{-2}}{2}\}$. It is easy to see that $\cal A$ partitions into two subsets, one containing the $2\times 2$ blocks along the diagonal and the other containing the off-diagonal blocks. Note that each of the sub-codes has two Alamouti blocks. With the aid of hard-limiting, the ML decoding complexity of this code can be reduced to $\sqrt{M}$ as in the case of the Srinath-Rajan code.
\end{example}

\section{Conclusion}
We have proposed a novel application of TOPS filters for STBCs and showed that the application of TOPS filters across space and time results in the partial removal of IAE interference. We have shown that any STBC which partitions into non-overlapping sets in space and time can be decoded independently with the aid of the TOPS filters. Furthermore, the reduction in the ML decoding complexity due to the application of TOPS filters in some of the best known codes in the literature has been studied. We have shown that the order of the ML decoding complexity reduces greatly in all the example STBCs considered.

\bibliographystyle{ieeetr}

\end{document}